\documentclass[amsmath,amssymb,preprint]{revtex4}
\usepackage{amssymb}
\usepackage[dvips]{graphicx}
\usepackage{color}

\begin{document}

\title{Electron-electron and electron-phonon relaxation in metals excited by optical pulse}
\author{V.V. Kabanov}

\affiliation{Department for Complex Matter, Jozef Stefan Institute, 1001 Ljubljana, Slovenia}

\begin{abstract}
A short overview of theoretical models for the description of the relaxation processes in metals excited by a short laser pulse is presented. The main effort is given to description of different processes which are taking place after absorption of the laser pulse. Widely used  two temperature model is discussed and the conditions of applicability of this model are identified. Various approaches for solving the Boltzmann kinetic equations are discussed. It is identified that in the case of low excitation limit the relaxation is determined by the emission of phonons by photoexcited electrons. The possibility to obtain the value of the electron phonon coupling constant from experiments  is discussed.
\end{abstract}

\pacs{71.38.-k, 74.40.+k, 72.15.Jf, 74.72.-h, 74.25.Fy}

\maketitle

More than 60 years ago the paper of M.l. Kaganov, I. M. Lifshitz and L. V. Tanatarov about relaxation between electrons and crystalline lattice in metals \cite{Kaganov} was published in Journal of Experimental and Theoretical Physics. The main motivation to consider this problem was the observation of the deviation from the Ohm's low in metals in the limit of strong current. As it was pointed by the authors in that case the energy accumulated in the electronic system differs considerably from that which corresponds to the lattice temperature $T$. In the paper it was postulated that the time required to establish equilibrium in the electron gas is much less than the time for achieving equilibrium between the electrons and the lattice, therefore the electron gas may be considered in a state of equilibrium, i.e., its state is described by the ordinary Fermi distribution function with certain nonequilibrium electron temperature $T_e$. This assumption allows to evaluate the rate of energy transfer between electrons and phonons.

Interesting extension of this work was proposed by V.A. Shklovskii \cite{shklovskii}.
He considered the size effect in heat transfer to the insulating substrate and in the nonlinear electric resistivity of current-heated metallic films deposited on a dielectric substrate. Using the same assumption that electrons have nonequilibrium temperature $T_e$ it was shown that the heat flux have qualitatively different behaviuor in the case of thick $d>>l_{pe}$ and thin $d<<l_{pe}$ film ($d$ is the thickness of the film and $l_{pe}$ is the phonon mean-free path due to scattering by electrons). In the first case the heat flux is determined by the transparency of the interface between metal and insulator, while in the former case the heat flux does not depend on the transparency of the interface between metal and insulator and determined only by the strength of electron-phonon intraction as described in Ref. \cite{Kaganov}. It is interesting to underline that concerning the definition of the electron temperature $T_e$ in Ref. \cite{shklovskii} it was stated that electronic temperature is determined by electron-electron collisions only if $T_e<k_B T_D^2/\epsilon_F$ where $T_D$ is the Debye temperature, $\epsilon_F$ is the Fermi energy and $k_B$ is the Boltzmann constant. This model was further developed in Refs. \cite{Bezuglyj1,Bezuglyj2,Bezuglyj3}. In particular, in these papers the nonequilibrium electron distribution function is described in terms of the electron temperature $T_e$. On the other hand phonons are described by a nonthermal distribution function \cite{Bezuglyj1,Bezuglyj2,Bezuglyj3}. Since the electronic temperature is not well defined, it was proposed some way to define this temperature from experiments \cite{Bezuglyj1,Bezuglyj2}. Note, that this theory is strongly based on the competition of the processes of electron-phonon thermalization and phonon escape from the film to the substrate. In the pump-probe experiments the observed relaxation processes are usually on the sub-picosecond time scale and are faster than any diffusion processes from the excitation volume.

Further generalization of the result of Ref. \cite{Kaganov} was proposed in Ref.  \cite{anisimov}. In this paper the results of Ref. \cite{Kaganov} were used in order to evaluate the emission current and the emitted charge from a metal surface exposed to a picosecond laser pulse. Here the idea of describing electrons and phonons as  quasiequilibrium states with nonequilibrium electronic $T_e$ and phonon $T_l$ temperatures was further developed. The system of two differential equations for $T_e$ and $T_l$ was derived. This model is called the two temperature model (TTM) and now widely used for the analysis of different experiments.

In 1987 Phillip Allen \cite{allen} pointed out, that optical pump-probe experiments may be used in order to determine the electron-phonon coupling constant $\lambda$ which plays a crucial role in the Eliashberg theory of superconductivity \cite{eliashberg2,eliashberg4}.
For that purpose the results of Refs. \cite{Kaganov,anisimov} were reformulated in terms of spectral function of the electron-phonon interaction $\alpha^2F(\omega)$, known as the Eliashberg function \cite{eliashberg2,eliashberg4}. As a result, measuring the relaxation time in pump probe experiments and comparing it with the predictions of TTM and Allen's theory provides direct information about some moments of the Eliashberg function
\begin{equation}
\lambda\langle\omega^n\rangle =2\int_0^{\infty}d\omega{\omega^n\alpha^2F(\omega)\over{\omega}}.
\end{equation}

The measurements were performed for a large number of known ordinary \cite{brorson} and high-$T_c$ superconductors \cite{chekalin}.

Note that Allen stated in his derivation of TTM, that the assumption that electrons and phonons have thermal distribution functions is at some level incorrect, probably in detail or possibly more seriously. Nevertheless, he pointed out that deviations from local thermal equilibrium may not in fact have much influence on the energy relaxation described by TTM.

Up to now, detailed experimental data on relaxation processes are available for metals \cite{brorson,schoenlein,Elsayed-Ali,groeneveld1,groeneveld2,brorson2,Elsayed-Ali2,Hartland}, high-temperature superconductors \cite{chekalin,eesley,han,albrecht,stevens,demsar,kabanov,gadermaier}, and pnictide superconductors \cite{stojchevska,rettig} using standard optical pump optical probe technique. Recently, a comprehensive analysis of experimental data on optical pump broad-band probe in high-temperature superconductors has been performed \cite{conte}. Most of the data are analyzed in the framework of the two-temperature model.

The TTM has well defined predictions as far as the temperature dependence of the relaxation time is concerned \cite{anisimov}. Very soon after successful determinations of $\lambda\langle\omega^2\rangle$ for different superconductors \cite{brorson,chekalin} the temperature and fluence dependences of the energy relaxation time for thin films of Ag and Au were obtained using the combined femtosecoud optical transient-reflection techniques with the surface plasmon polariton resonance \cite{groeneveld1,groeneveld2}. The results clearly indicated that the temperature and laser intensity dependence of the energy relaxation rate is in clear contradiction with the TTM (see Fig.1 taken from Ref. \cite{groeneveld2}). Experimentally a decrease of the effective relaxation time for both Ag and Au thin films was observed when the temperature is lowered from 300 to l0 K. The relaxation time did not show any increase in the low temperature limit. Also the relaxation time does not show any dependence on the deposited laser energy density \cite{groeneveld1,groeneveld2}. In order to resolve discrepancies the electron-electron collisions were included in the Boltzmann kinetic equations. Numerical integration of these nonlinear equations gives satisfactory description of the observed relaxation time \cite{groeneveld1,groeneveld2}.

\begin{figure}
\includegraphics[width = 90mm, angle=0]{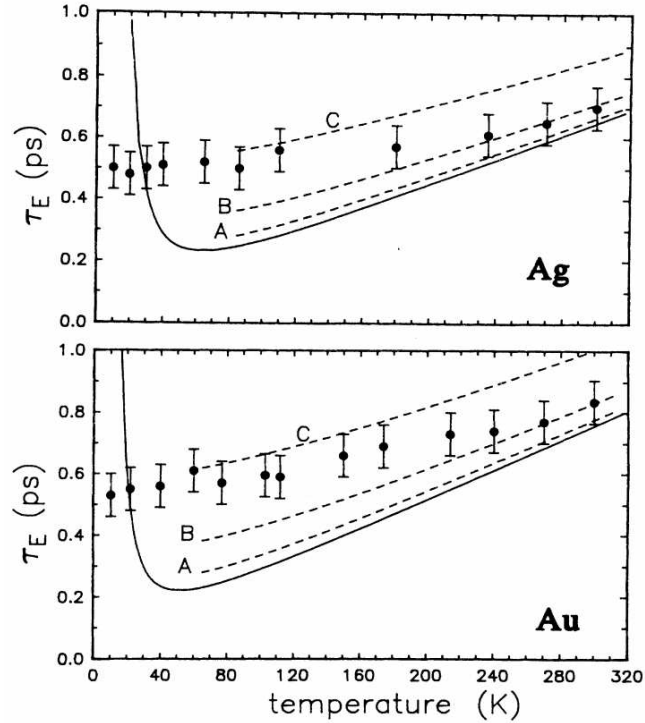}
\caption{Comparison between experimental data and the calculated electron-phonon energy relaxation time $\tau_E$ in the TTM versus temperature for various deposited laser-energy density $U_l$ for Ag and Au. Solid line: infinitesimally small $U_l$. Dashed lines A, B, C: $U_l$ =0.3, 1.3, 5.2 J cm$^{-3}$, respectively. Dots: experimental data.}
\end{figure}

At this point we have to remind about applicability of the Boltzmann kinetic equations to photo-excited metals. According to Ref. \cite{PhysKinet} the characteristic distances where the distribution function changes should be larger than $1/k_F$ where $k_F$ is the Fermi momentum. Usually for pump-probe experiments this is not a problem. If the experiments are performed on thin films with the thickness less than the absorption length this is not a problem at all. In the case of bulk materials we have to compare the absorption length with $1/k_F$. Usually the absorption length in metals is of the order of ten nanometers or more which allows to use kinetic approach. Another restriction is related to the characteristic time of the evolution of the distribution function. This time must be longer that $\hbar/\epsilon_F\sim$ 1fs, where $\hbar$ is the Planck constant. In most of the pump probe experiments in metals the time resolution is limited by tens of femtoseconds. It also allows to use the Boltzmann kinetic equations. More attention should be paid when the distribution function has high energy tails ($\epsilon >\epsilon_F$) which may relax very quickly.

In order to clarify the situation let us make some estimates and clarify the regions of applicability of different theories. In accordance with Ref. \cite{kabanovAlex} the thermalization of electron gas due to electron-electron collisions occurs at the time scale $\tau_{ee}=4\hbar\epsilon_F/\pi^3\mu_c^2(k_BT)^2$, here we introduce Coulomb pseudopotential $\mu_c$. The Coulomb pseudopotential plays also very important role in the Eliashberg theory of superconductivity \cite{eliashberg2,eliashberg4}. Note, that in the superconductivity theory the Coulomb pseudopotential is reduced due to retardation effects \cite{MorelAnderson,Tolmachev}. Electron-phonon thermalization occurs on the time scale determined by TTM \cite{anisimov,allen}. At low temperatures, $T<T_D$, $\tau_{e-ph} =\pi\hbar^3\omega_D^2/720\zeta(3)\lambda(k_BT)^3$. Here $\lambda$ is determined by Eq.(1) with $n=0$ and $\omega_D=k_BT_D/\hbar$ is the Debye frequency and $\zeta(3)\approx 1.202$ is Riemann zeta function. At high temperatures $T>T_D$ the electron-phonon thermalization is determined by formula \cite{allen,kabanovAlex}, $\tau_{e-ph} =\pi k_BT/3\hbar^3\lambda\langle\omega^2\rangle$. The TTM is justified if the thermalization due to electron-electron collisions is faster than electron-phonon thermalization $\tau_{e-e}<\tau_{e-ph}$. Assuming that the dimensionless parameters $\mu_c$ and $\lambda$ are of the order of 1 and dropping all numerical coefficients we obtain that TTM is applicable in the range of temperatures \cite{baranov}: $T_D (\hbar\omega_D/\epsilon_F)<T< T_D (\epsilon_F/\hbar\omega_D)^{1/3}$. This inequality means that the TTM is not applicable in the temperature range where most of the pump-probe experiments are performed.

There is an additional problem of the TTM which is related to the fluence dependence of the pump-probe experiments. Experimentally it is relatively easy to reach such intensity of laser pulses that the electronic temperature is much higher than the lattice temperature $T_e \gg T$. It means that in accordance with the TTM the response must be a nonlinear function of the pump power at such intensities. This was not reported in Refs. \cite{groeneveld1,groeneveld2}. In order to see the nonlinearities in the pump probe experiments it is necessary to use much higher laser intensities as in experiments  \cite{groeneveld1,groeneveld2} that the calculated electronic temperature reaches  $T_e=T_D (\epsilon_F/\hbar\omega_D)^{1/3}$ limit. In experiments \cite{groeneveld1,groeneveld2} the laser intensities was high enough that the calculated electronic temperature satisfies the first inequality $T_e \gg T$. On the other hand, the estimated electronic temperature was not large enough to observe nonlinearities $T_e\ll T_D (\epsilon_F/\hbar\omega_D)^{1/3}$. There are reports of the pump-probe measurements in metals at very high fluences when the estimated electronic temperature exceeds the limit $T_D (\epsilon_F/\hbar\omega_D)^{1/3}$ \cite{Elsayed-Ali2,Hartland}. As expected, in that case the TTM is applicable and the fluence dependence of the relaxation time is observed. Note, that the TTM was successfully used for the interpretation of the photo-induced suppression of the spin density wave in chromium \cite{Nicholson}. Similarly, the TTM provides reasonable description of the nonequilibrium dynamics in the topological insulator Bi$_2$Se$_3$ \cite{Sobota}. In both cases the pump intensity was high enough to rise the electronic temperature above the limit  $T_D (\epsilon_F/\hbar\omega_D)^{1/3}$ \cite{Nicholson,Sobota}.

These arguments suggest that during these type of experiments the electron distribution function is far from equilibrium and cannot be described by the quasiequilibrium Fermi distribution function with elevated electronic temperature. Most convincing arguments against the TTM were obtained by  the direct measurements of the electron distribution function using time-resolved photoemission spectroscopy \cite{fann,lisowski,perfetti}. The
interpretation of the photoemission spectra in terms of the distribution function of electrons assumes that the matrix element involved in the photoemission experiments and the density of electronic states near $\epsilon_F$ are smooth functions of energy \cite{reinert}. In the case of ordinary metals, this assumption is usually correct. In some metals, additional bands appear in the spectra, but usually these bands have relatively large excitation energy and do not influence the determination of the energy dependence of the distribution function (see Fig. 5 in Ref. \cite{lisowski}, for example). In the measurements \cite{fann,lisowski,perfetti}, the transient electron distribution function is not thermal and has high-energy tails, which survive till the thermalization
occurs \cite{fann}. Moreover, in gold at about 400 fs after excitation, about 30\% of the pump energy is already in the phonon subsystem \cite{fann}, while the thermalization is observed only after 1 ps. Therefore the transfer of energy from electrons to
phonons occurs faster than the electronic thermalization. A similar effect is observed in Ruthenium \cite{lisowski}. It is estimated that 100 fs after the pump about 20\% of quasiparticles are in the high energy tails of the distribution function. To account for this effect it was suggested that the electron distribution function can be represented as a sum of thermal and nonthermal parts \cite{fann,lisowski}. It was also suggested to approximate the nonthermal distribution function by a Fermi-Dirac function with reduced amplitude and a nonphysical auxiliary temperature \cite{lisowski}. Therefore, the main conclusion of the time resolved photoemission experiments is that significant part of the energy is transferred from electrons to phonons before electrons are thermalized. It demonstrates the direct contradiction with the assumptions of the two temperature model.

In high temperature superconductors time resolved photoemission experiments also show that high energy tails in the electron distribution function survives until few hundreds of femtoseconds \cite{perfetti}. More sophisticated analysis of the time resolved ARPES spectra of Bi$_2$Sr$_2$CaCu$_2$O$_{8+x}$ is presented recently in Ref. \cite{Rameau}. Measuring relaxation time for photo excited electrons at different energies the different relaxation processes were identified. At high energies the relaxation is governed by fast electron-electron collisions. At at lower energies the population decay is governed by optical phonons with the time scale in the range from tens to hundreds of
femtoseconds. And finally, the thermalization occurs on the picosecond time scale \cite{Rameau}. The results of time resolved APES data were compared with simulations which include electron-electron and electron-phonon collisions and are based on quantum kinetic equations for non-equilibrium Keldysh Green’s functions \cite{Kemper,Sentef}. To describe the experimental observations the Eliashberg function was modeled by two delta functions at energies corresponding to acoustic and optical phonons \cite{Rameau}. Overall, the qualitative agreement of experimental data and simulations was quite good.

These experimental facts make the application of the TTM questionable and the results obtained on the basis of the analysis of experimental data applying the TTM unreliable. Understanding of this fact initiated many publications where different approaches were used in order to simulate the Boltzmann kinetic equations for electrons and phonons in metals excited by ultrashort laser pulses \cite{lugovskoy,rethfeld}. For example in Ref. \cite{lugovskoy} the electron-phonon collision integrals were considerd exactly while the electron-electron collisions were considered in the relaxation time approximations. As a result, the evolution of the calculated electron distribution function is similar to that observed in experiments in Ref. \cite{fann}. On the other hand in Ref. \cite{rethfeld} the complete set of kinetic equations was simulated numerically. For laser excitations near the damage threshold it was found that the energy exchange between electrons and lattice can be described with the two-temperature model, in spite of the nonequilibrium distribution function of the electron gas. In contrast, the nonequilibrium distribution leads at low excitations to a delayed cooling of the electron gas, indicating the deviations from the TTM. The conclusion was that the cooling time of laser-heated electron gas depends on excitation parameters.

The description of the relaxation of hot electrons in metals with a numerical solution of the Boltzmann kinetic equations demonstrates qualitative agreement with experiments. The main problem of this description is the lack of estimates of observable time scales as well as the lack of description of different physical processes which occur in a metal on a short time scale. In Ref. \cite{gusevWright} qualitative arguments were presented in order to analyze the processes in metal after photo-excitation. According to this theory at the first stage of relaxation multiplication of the electron-hole pairs due to electron-electron collisions occurs. The quasiparticle density increases in accordance with the scaling low $n\propto\sqrt{t}$ \cite{gusevWright}. On the other hand, each quasiparticle emits phonons  with the rate $\pi\lambda\langle\omega\rangle$. This leads to the transfer of energy to phonons in accordance with the formula $E\propto 1-(t/\tau_E)^{3/2}$, where $\tau_E=4^{1/3}\epsilon_F^{1/3}/\pi^{4/3}\hbar^{1/3}(\mu_c\lambda\langle\omega^2\rangle)^{2/3}$. The value of $\tau_E$ was estimated as 0.8 ps in gold. Note, that the relaxation time $\tau_E$ is defined as if at $t=\tau_E$ all energy absorbed by electrons is transferred to phonons in agreement with photoemission experiments \cite{fann}. 
Here we have to note that relaxation time measured by pump-probe technique shows linear temperature dependence and is of the order of 1 ps. Therefore, the estimate for $\tau_E$ is not consistent with the experiments.

Later on in Ref. \cite{kabanovAlex} the linearized Boltzmann kinetic equations for the distribution function averaged over angles were formulated. Both electron-electron and electron-phonon collision integrals were considered. For electron-electron collisions the integro-differential kinetic equation is reduced to differential form for the Fourier transformed distribution function \cite{kabanovAlex}. This equation is similar to the diffusion equation and contains only one characteristic time scale $\tau_{ee}=4\hbar\epsilon_F/\pi^3\mu_c^2(k_BT)^2$, which plays the role of diffusion coefficient in this equation. The relaxation is determined by the spectrum of eigen values of the differential operator defined in Ref. \cite{kabanovAlex}. This differential operator is similar to a one-dimensional quantum mechanical Hamiltonian \cite{kabanovAlex}. The smallest eigen value in the continuous spectrum of this operator defines the longest relaxation time which is exactly equal to $\tau_{ee}$. This timescale determines the electron-electron thermalization. Initial fast electron-electron relaxation is determined by the large eigen values of this operator. The initial fast relaxation strongly depends on the initial distribution function (namely on the frequency of the laser pulse). But, irrespective of the initial excitation, the asymptotic behaviour on the large time scale is determined by $\tau_{ee}$:
\[
n-n_{EQ} \propto \frac{\exp{(-t/\tau_{ee})}}{t}
\]
where $n_{QE}(\epsilon)=\frac{(T_e-T)}{4T}\frac{\epsilon/k_BT}{\cosh{(\epsilon/k_BT)}}$ is the thermal distribution function described by the nonequilibrium electronic temperature $T_e$ in the limit $T_e-T\ll T$.

For the electron-phonon collision integrals two limiting cases were proposed. In the low temperature limit $T<T_D$ and in the case of Eliashberg function for "dirty" metals $\alpha^2F(\omega)=\lambda\omega/(2\omega_D)$ for $\omega<\omega_D$, the Boltzmann kinetic equation is reduced to a differential form, similar to that for electron-electron collisions \cite{kabanovAlex}. In this limit there is only one timescale, determined by this equation $\tau_{e-ph}=\frac{2\hbar^2\omega_D}{\pi^3\lambda(k_BT)^2}$ which also plays the role of diffusion coefficient in the equation. Since for any realistic case $\tau_{ee}\gg\tau_{e-ph}$ the limit of the TTM is never achieved. As in the case of electron-electron collisions the thermalization at low temperatures is determined by the smallest eigen value in the continuous spectrum of the differential operator defined in \cite{kabanovAlex}. Thermalization time is equal to $\tau_{e-ph}$.  The asymptotic at large time has the following form:
\[
n \propto \frac{\exp{(-t/\tau_{e-ph})}}{t}
\]
Therefore, experimentally only $\tau_{e-ph}$ is achievable. As in the case of electron-electron collisions the relaxation on short time scale is not universal and is determined by the details of the initial distribution function.

The electron-phonon collision integrals may be rewritten in the differential form in the limit of high temperature $T>T_D$ or in the limit when the average energy per excited quasiparticle is larger than $\hbar \omega_D$ \cite{kabanovAlex,baranov}. In that case the equation for the nonequilibrium part of the distribution function $\phi(\epsilon,t)$ has the form:
\begin{equation}
\gamma^{-1}\dot{\phi}(\epsilon,t)=\frac{\partial}{\partial \epsilon}\Bigl(\tanh{(\epsilon/2k_BT)}\phi(\epsilon,t)+k_BT\frac{\partial}{\partial \epsilon}\phi(\epsilon,t)\Bigr),
\end{equation}
where $\phi(\epsilon,t)=f(\epsilon,t)-f_0(\epsilon,T)$. Here $f(\epsilon,t)$ is the nonequilibrium distribution function for electrons and $f_0(\epsilon,T)$ is the equilibrium Fermi-Dirac  distribution function at equilibrium temperature $T$ and $\gamma=\pi\hbar\lambda\langle\omega^2\rangle$. This equation shows that the electron-phonon relaxation represents the diffusion of electrons in energy space with emission of phonons. It is physically justified, because the characteristic energy of electron is higher than the phonon energy.

In order to apply this equation for description of the relaxation photo-excited electrons after a short laser pulse it is necessary to add linearized electron-electron collision integral to the right hand side of equation (2)\cite{baranov}. Experimentally it is known \cite{fann,lisowski,perfetti} that during subpicosecond relaxation the average energy per photoexcited electron is relatively large. Therefore, Eq.(2) with the electron-electron collision integral may be applicable for the description of experimental data. Indeed the numerical simulations performed in Ref. \cite{baranov} showed that equation (2) compared to exact electron-phonon collision integrals with different Eliashberg functions describes the evolution of the distribution function well enough in the whole process of relaxation. In Fig.2 we presented the comparison of relaxation times obtained from the simulation of the linearized Boltzmann kinetic equations and the Fokker-Planck equation (2). The results demonstrate that Eq.(2) describes the relaxation quite well.

\begin{figure}
\includegraphics[width = 90mm, angle=0]{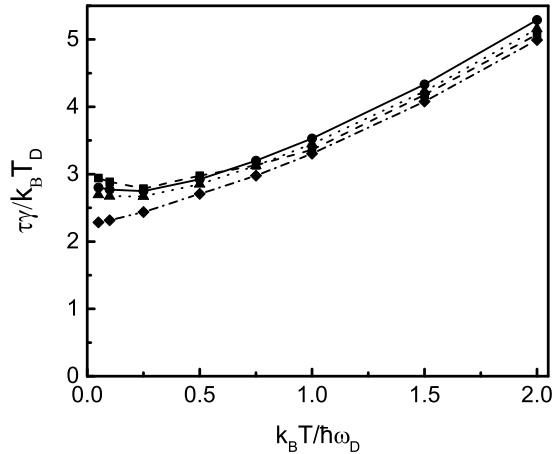}
\caption{Temperature dependence of the energy relaxation time calculated for three different Eliashberg functions. Dotted line with triangles represents the "dirty metal case" $\alpha^2F(\omega)=\lambda\omega/2\omega_D$, solid line with circles represents Debye model $\alpha^2F(\omega)=\lambda\omega^2/\omega_D^2$, and dashed line with squares represents the Einstein model $\alpha^2F(\omega)=\lambda\omega_0\delta(\omega-\omega_0)$. Dashed-dotted line with diamonds represents temperature dependence of the relaxation time for the Fokker-Planck equation (2). In the case of Einstein model, $T_D = \omega_0/k_B$.}
\end{figure}

Very often there is no need to solve the full set of kinetic equations. Sometimes it is sufficient to describe the energy transfer from electrons to phonons. For this purposes we can multiply Eq.(2) by the energy and integrate over $\epsilon$. The integral which involves the electron-electron collision integral reduces to 0 because electron-electron collisions preserve the total energy. The integration yields:
\begin{equation}
\dot{E}=-2\gamma\int_0^\infty d\epsilon\tanh{(\epsilon/2k_BT)}\phi(\epsilon,t).
\end{equation}
This equation describes the emission of phonons by nonequilibrium electron-hole pairs.
If we assume that the characteristic energy scale in $\phi(\epsilon,t)$ is large in comparison to $k_BT$ the equation is reduced to equation $\dot{E}=-2\gamma n$ derived in Ref. \cite{gusevWright}, where $n$ is the nonequilibrium density of electron-hole pairs. It has very straight forward interpretation. The high energy electron-hole pairs emits phonons with the rate
$\tau_{em}^{-1}=\pi\lambda\langle\omega^2\rangle/\langle\omega\rangle$.

Therefore, we have the following physical interpretation of the relaxation process after photoexcitation \cite{baranov}. The pump pulse creates a broad distribution of electron-hole pairs with large excitation energy. The high-energy electrons relax to the low-energy scale $\epsilon \sim (\hbar\omega_D\epsilon_F)^{1/2}\gg k_BT,\hbar\omega_D$ due to electron-electron collisions. The energy scale is determined by the condition when the life time of electron due to electron-electron collisions becomes equal to the life time due to electron-phonon collisions. For electrons with energy $\epsilon < (\hbar\omega_D\epsilon_F)^{1/2}$ the electron-electron collisions may be neglected. It happens on the time scale $\langle\omega\rangle/\lambda\langle\omega^2\rangle$. The photoexited electron-hole pairs emit phonons immediately after excitation. The emission rate is temperature independent and is not affected by the Fermi distribution function, because the average energy of nonequilibrium electrons is large in comparison to $k_BT$ and the distribution function for electrons is small at that energy. The solution of Eq.(2) shows that the relaxation time is weakly temperature dependent in agreements with experiment. The divergence of the relaxation time at low temperatures is absent \cite{baranov}. It does not depend on the laser pump fluence and laser frequency. Eq.(2) allows the simple estimate of the relaxation time. Since the characteristic energy of nonequilibrium electrons is of the order $(\hbar\omega_D\epsilon_F)^{1/2}$ the derivative of the distribution function may be estimated as $\partial \phi/\partial \epsilon \sim \phi/(\hbar\omega_D\epsilon_F)^{1/2}$. Therefore, the relaxation time determined by Eq.(2) may be estimated as $\tau\approx (\hbar\omega_D\epsilon_F)^{1/2}/\pi\hbar\lambda\langle\omega^2\rangle \approx 1$ ps for gold.

Note that electron-electron collisions have an important effect on energy relaxation. When the electron-electron collision rate increases, the high-energy excitations are decaying with the creation of more low-energy electron-hole pairs. The energy relaxation on the other hand is proportional to the number of nonequilibrium electrons. Therefore when the electron-electron collision rate increases the energy relaxation rate of nonequilibrium electrons increases as well. This relaxation occurs until almost all energy is transferred to phonons. Only after that the thermalization, described by the set of equations derived in Ref. \cite{kabanovAlex}, starts. But, since most of the energy is transferred to the phonon subsystem before the time when the average energy per electron-hole pair becomes of the order of $k_BT$, the final and the longest stage of thermalization is difficult
to observe.

Note, that recently it was suggested \cite{Waldecker,Maldonado}, that phonon distribution function is not thermal and it plays an important role in relaxation of photo excited electrons. In the weak perturbation limit it was shown that the phonon distribution function is always thermal and may be described by the nonequilibrium themperature \cite{baranov}. In the case of strong perturbation, when electronic temperature is more than 1000K, it appears that these corrections are important \cite{Maldonado}, and phonon distribution function remains unthermal on the time scale of few tenth of picoseconds. The relaxation of nonequilibrium phonons was also studied in recent publication \cite{Klett}. In addition to electron-electron and electron-phonon collisions the phonon-phonon collisions were considered. The main conclusion of these calculations is that the relaxation of phonons occurs on the much longer time scale than the relaxation of hot electrons. However, the main question remains if this relaxation may be observed experimentally.

In conclusion it is important to underline that the solution of the Boltzmann kinetic equations provides good description of the hot electron relaxation in metals excited by the short laser pulse. But because the relaxation time depends not only on the electron-phonon coupling constant $\lambda$, but, also depends on the electron-electron Coulomb pseudopotential $\mu_c$ the unique way to determine $\lambda$ is not possible. Nevertheless, using high power laser pulses it is possible to reach very high electronic temperatures $T_e > T_D(\epsilon_F/\hbar\omega_D)^{1/3}$ which allows the application of the TTM. In that case the coupling constants may be determined in the standard way \cite{demsar2}. However, the applicability of the TTM must be verified via measurements of temperature and fluence dependence of the relaxation time \cite{demsar2}.

The author acknowledges the financial support from the Slovenian Research Agency Program No. P1-0040.

\end{document}